# Thin film optics computations in a high-level programming language environment: tutorial

SALVADOR BOSCH,[1,*] JOSEP FERRÉ-BORRULL,[2] JORDI SANCHO-PARRAMON[3]

[1]Universitat de Barcelona, Dep. Física Aplicada, Martí i Franqués 1, 08028 Barcelona
[2]Universitat Rovira I Virgili, Eng. Electrònica, Elèctrica i Automàtica, Campus Sescelades, Avinguda Països Catalans, 26, 43007 Tarragona
[3]Ruder Bosckovic Institute, Laboratory for optics and optical thin films, Bijenicka cesta 54, 10000 Zagreb
*Corresponding author: sbosch@ub.edu

**Thin film technology is a most relevant field in terms of number of different applications and, even more, in how widespread the use of the technology is. Virtually all optical devices involving light beams or with imaging capabilities available on the market contain thin film assemblies that are crucial for enhancing the practical performances of the device. So, nowadays, thin film coatings are routinely deposited on very different kinds of substrates. From the beginning, the use of thin films required the optical characterization of the coating materials as well as the design of the suitable thin film structures in order to meet the optical requirements. This has been a significant research topic in the past but the impact of high level programming languages has dramatically changed the scenario. Nowadays, to address the more advanced computation tasks related to thin film technology is much simple. The aim of the Tutorial presented here is to enable the reader (with only a basic knowledge of the use of Matlab or Python) to perform advanced numerical computations of the optical properties of thin film structures. We will explain the necessary steps and supply the corresponding program codes, with complete comments. For conciseness, our short theoretical introduction will follow the scheme established elsewhere and we will concentrate on developing in detail several practical examples of increasing difficulty, so that understanding and using the Tutorial becomes entirely intuitive.**

## 1. INTRODUCTION

Thin film coatings can be found in a wide variety of devices and are truly ubiquitous in the daily life of billions of people. The technology has become so ingrained in our global society that it is often overlooked. For example, in smartphones thin films have a profound effect to improve touch features, increase the transmittance of touch panels, and lengthen the lifespan of displays. For liquid crystal displays applications, optical thin film coatings are used in regard to these devices to produce lenses with virtually no reflection, metal films with a high level of reflectance, and optical filters and mirrors that are used to create rich, vibrant colors. Reflecting telescopes utilize optical thin films to produce mirrors that are highly reflective. In solar energy, solar panels performance rely on, for example, high-durability transparent thin films that protect their surface. In transportation, defrosters and transparent heaters that are used to keep the windows of vehicles from freezing during cold weather make also use of thin film technology.

The computation of the optical performances (reflectance, transmittance, phase changes upon light reflection, etc...) of thin film layered structures is a topic of major practical interest in all the mentioned areas. In this context, the aim of this Tutorial is to provide a set of tools and methodology that make numerical computations of the optical properties of thin film structures easily accessible to interested researchers. We choose a widely spread high-level programming language environment such as Matlab or Python to implement these tools. The appropriate theoretical framework for the problem is the electromagnetic theory of light and a good knowledge of this theory is required for addressing all the related technological issues in depth. The basic underlying assumption is that light interference governs the pass of light through the layers, because of its 'thin' character. This normally means physical thicknesses up to some microns. Conversely, we will always assume that the pass of the light through the substrates is described in terms of beam intensities. All these aspects, for our present development, are established in detail in the references significant in the field [1–8] and, for clarity and conciseness, we adopt as a reference text for our approach the optical theory as it is developed in ch. 2 of Macleod's book [2]. This textbook is one of the most renowned in the field from its first edition and has updated its contents continuously. Besides, not only theoretical aspects are addressed there but also many practical topics (such as coating the two sides of a substrate) as well. Thus, basing our developments mostly in this single reference greatly simplifies our presentation and also facilitates a deeper study to the interested developer. We will assume that the reader has an essential knowledge of that text, since it is a requirement for our present work. A deep and detailed understanding is not required, as far as all the definitions and formulas presented therein are clearly understood. Operating this way we ensure a sound theoretical basis for our work by simply including here a short summary of the important theoretical results and, at the same time, using for our program codes notations similar to our reference text.

It is important to note that in the present Tutorial we will address the general configuration case: thin film structures with any number of layers, with transparent or absorbing materials, with oblique incidence angle and incident polarized light beams. Thus, we think that the utility of our present work is very wide. We do not allow anisotropy in the materials; this is not a serious limitation in practice, since this is a very advanced and specific topic of very specialized practical interest.

In short, a thin film structure consists of a series of plano-parallel layers alternating different materials (usually transparent) deposited on a substrate. When light incides on the structure, the interference effects originated at the interfaces of the layers give rise to some resulting reflected and transmitted beams. Overall, the properties of the reflected and transmitted beams can be dramatically changed depending on the combination of refractive indices and thicknesses of the layers of the structure. In the most general case, the refractive indices are dispersive (vary with the wavelength of the incident beam) and able to represent absorbing materials, by being complex numbers with a non-zero imaginary part. The incidence angle of the light beam is also a characteristic parameter for the problem. Thus, the primary objective of a computing program for thin films is to determine the characteristics of the resulting reflected and transmitted beams, from the knowledge of the incident beam and the design parameters of the structure.

This brief description is useful to explain our Tutorial as a practical approach to build a computation program. After a concise theoretical introduction where we simultaneously introduce the designation of the magnitudes in our codes, we explain the precise optical description of the thin film structure. Next, the electromagnetic theory is used to physically describe the light interference effects at all the interfaces, leading to some convenient key parameters (basically, the overall reflectance and transmittance coefficients) to characterize the structure. The core of the theory is built inside some numerical 'functions' that the user does not need to change. Finally, once the overall parameters mentioned are calculated, the last part of the code is to use them for finding the relevant measurable magnitudes for the problem under study (say light intensity reflectance, transmittance, etc). A first basic example illustrates the simplicity of the overall procedure. In a second example (more advanced) we will show all the relevant programing details. The third example illustrates how our computational procedures are able to address specialized research topics by simple adaptations of our codes. The text develops the three examples giving all the explanations on how to write the code. The full codes for running them are presented all together in the supplementary material for better clarity. Further details related to how to calculate other interesting quantities, explanations about some functions and how to enhance the capabilities of the present software are developed in the Appendix.

For conciseness, the examples are presented in Matlab language, where a 'function' is some limited code that works as a "black box" object in which the user provides a set of input arguments and the function returns some output. A 'script' in Matlab is a general piece of code that handles the operations and calls the functions and it is where the user should define the specific physical problem under consideration. It is also possible to include the functions inside the code of a script in a single file (we will do it in part 6 of the Supplementary material). The Python version of the codes is also included in the Supplementary material. In order to make the computational approach easily accessible to a wide range of researchers, we have prioritized clarity and simplicity rather than code succinctness or computational efficiency. Thus, only a basic knowledge of Matlab or Python is required in order to understand and easily generalize the presented examples.

## 2. SUMMARY OF THEORETICAL RESULTS

### A. Definition of the configuration

Thin film coatings are structures composed by plano-parallel layers made of materials that have different optical properties. Due to the interference effects generated by the assembly, dramatic changes in the optical performance of the setup can be achieved. The precise theoretical framework to approach the problem is the electromagnetic theory of light [1–8] and, apart from manufacture difficulties, a deep knowledge of this theory allows addressing all the design issues adequately.

In our present development, we adopt as a reference text for our approach, the optical theory as it is developed and presented in chapter 2 of Macleod's book [2]. We will summarize now the main results presented therein at the same time we introduce the names of the different magnitudes in our codes in accordance with the text. All our procedures will be valid for dispersive materials but the description here is monochromatic for simplicity.

Each material constituting any layer is isotropic and it has a definite complex refractive index ($N$) with the real part being the refractive index ($n$) and the imaginary part the extinction coefficient ($k$). For transparent materials the complex refractive index is purely real and accounts for the ratio between the speed of light in vacuum and in the medium (a real number). When the material is absorbing, the refractive index is a complex number with its imaginary part being the extinction coefficient, which describes the attenuation of the light beam as it propagates in the material [1,8]. We will number the series of materials increasing from the ambient (0) to the substrate (1, 2, ..., m), writing their complex refractive indices as $N_0, N_1, ... N_m$. In accordance, in this initial explanation the physical thicknesses are both 'infinite' for the ambient and substrate ($d_0 = d_m = \infty$) and $d_1, d_2, ... d_{m-1}$ for the layers (having units of length).

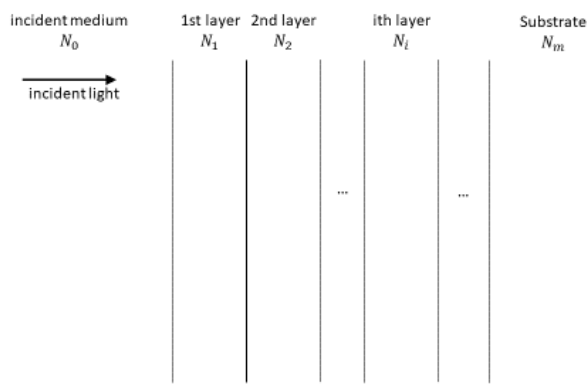

Figure 1. Layout of the basic physical configuration we consider here.

We will always realistically assume the 'ambient' where we perform the experiments to be transparent ($N_0 \in \mathbb{R}$) but we allow the rest of the materials to be absorbing, writing the complex refractive index for material of layer 'r' in the form $N_r = n_r - ik_r$, where 'i' is the imaginary unit.

The substrate may be coated also at the rear surface. If its shape is plano-parallel and its constitutive material is non-absorbing, later we will explain how to assume $d_m \neq \infty$ and calculate transmittances and reflectances also for this case.

There is only one normal direction to all the interfaces. If this normal coincides with the incident light beam direction, we have the normal incidence case. If not, the oblique beam direction and the normal direction define the incidence plane in our problem, being the incidence angle named $\vartheta_0$. Following our reference text, in our developments we will consider only the components of the electric and magnetic fields that are parallel to the interfaces [2]. Besides, it is important to note that the electromagnetic theory of light demonstrates the validity of the Snell law for complex refractive indices for the rays trough the layers [2]. Thus, the following expression is valid for the directions of one incident ray (form medium $N_0$ with angle $\vartheta_0$) through the layers of the structure:

$$N_0 \sin \vartheta_0 = N_1 \sin \vartheta_1 = ... = N_m \sin \vartheta_m \quad (1).$$

To develop the theory for the general case of oblique incidence with isotropic materials, the transverse electric field of the incident beam must be separated into its components parallel ('p', also known as 'TM') and the perpendicular ('s', also known as 'TE') to the incidence plane. This separation is crucial in all the theory below, since the two components do not mix or interfere and, simply, lead to two independent problems to be solved: the 'p' and the 's' problems. Moreover, the 's' components of all the fields are mutually parallel and inherently tangential to all the interfaces unlike the 'p' components, that change direction at each light refraction. However, it is usual to consider only the tangential parts of the 'p' electric fields, since they are the only responsible for energy fluxes in the direction normal to the layers. Thus, in what follows, two independent problems have to be solved separately: the 'p' problem (for its tangential part of the 'p' electric field) and the 's' problem, for this component of the electric field. The solving formalisms are quite similar, but not identical.

**B. Calculation methodology of the problem**

For each of the materials in our structure, assuming incidence angle $\vartheta_0$, let us introduce the tilted optical admittances [2] for the 'p' and 's' polarizations according to the formulas

$$\eta_p = \frac{N}{\cos \vartheta}, \quad \eta_s = N \cos \vartheta, \quad (2)$$

where expression (1) has to be used for computing $\vartheta$ inside the material. For each polarization, the numerical solving procedures use a 2x2 matrix for each layer. If the geometrical thickness of the layer is $d$, the corresponding matrix is

$$M = \begin{pmatrix} \cos \delta & (i \sin \delta)/\eta \\ i\eta \sin \delta & \cos \delta \end{pmatrix}, \quad \delta = \frac{2\pi N d \cos \vartheta}{\lambda}. \quad (3)$$

For the assembly, with our notation, we finally have

$$\begin{pmatrix} B \\ C \end{pmatrix} = M_1 M_2 ... M_{m-1} \begin{pmatrix} 1 \\ \eta_m \end{pmatrix}, \quad (4)$$

from where we compute $r = \dfrac{\eta_0 B - C}{\eta_0 B + C}$, $T = \dfrac{4 \eta_0 \operatorname{Re}(\eta_m)}{|\eta_0 B + C|^2}$. (5)

Since there are the two independent 'p' and 's' polarizations, with independent optical admittances as given by (2), we finally may define the four quantities

$$r_p, T_p, r_s, T_s. \quad (6)$$

Accordingly, in the code written for our developments, we have used the following names for these four most basic quantities: `rp, Tp, rs, Ts`.

The 'r' quantities are complex ratios for the electric field (between reflected and incident amplitudes). Then, the intensity reflectances are, simply,

$$R_p = |r_p|^2, \quad R_s = |r_s|^2 \quad (7)$$

and the phase changes upon reflection are usually defined by the single complex quantity ρ or, equivalently, by the two real quantities $(\Delta, \Psi)$ according to the following equations [9]

$$\rho \equiv \frac{r_p}{r_s} \equiv \tan \psi \cdot e^{i\Delta}, \text{ or by } (\tan \psi, \cos \Delta). \quad (8)$$

For the ratios between transmitted and incident beam intensities, the relevant quantities are, directly, $T_p$ and $T_s$ since usually only light intensity ratios are of practical interest for transmitted light.

When the substrate is transparent and coated at both sides (designed here 'a' and 'b') the relevant formulas for the overall intensity ratios are the following, according to the notations introduced in our Figure 2.

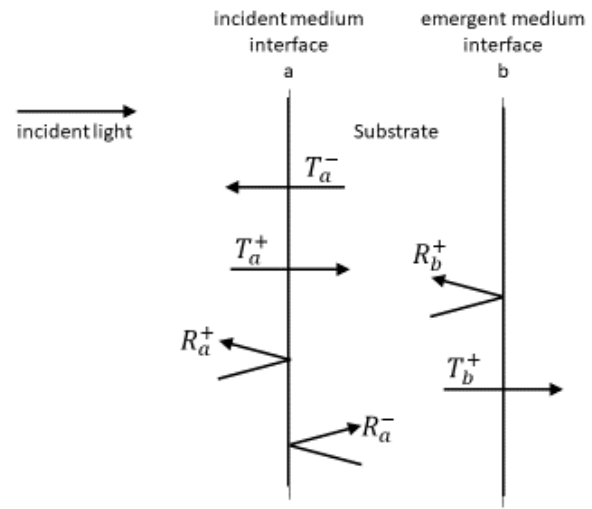

Figure 2. Explaining the notations used in the text for the case of a substrate coated on both sides (Equations 9 and 10).

$$T = \frac{T_a T_b}{1 - R_a^- R_b^+}, \quad R = \frac{R_a^+ + R_b^+ \left(T_a^2 - R_a^- R_b^+\right)}{1 - R_a^- R_b^+} \quad (9)$$

Note that when the substrate is not semi-infinite (i.e., when there is back side light reflection mixed with the front side reflected beam) it does not make sense to measure phase changes at reflection and only light intensity ratios are of practical interest.

Finally, considering both light polarizations 'p' and 's', all the relevant quantities at front and rear surfaces are:

$$R_p^{a+} = \left|r_p^{a+}\right|^2, R_p^{a-} = \left|r_p^{a-}\right|^2, T_p^{a+}, T_p^{a-}, R_p^{b+} = \left|r_p^{b+}\right|^2, T_p^{b+},$$
$$R_s^{a+} = \left|r_s^{a+}\right|^2, R_s^{a-} = \left|r_s^{a-}\right|^2, T_s^{a+}, T_s^{a-}, R_s^{b+} = \left|r_s^{b+}\right|^2, T_s^{b+} \quad (10)$$

In the code written for this tutorial, we have used the following names for these quantities:
Rp_aPlus, Rp_aMinus, Tp_aPlus, Tp_aMinus, Rp_bPlus, Tp_bPlus, Rs_aPlus, Rs_aMinus, Ts_aPlus, Ts_aMinus, Rs_bPlus, Ts_bPlus.

Accordingly, for the whole sample, the reflectances and transmittances obtained by using (9) will be designed: Rp, Rs, Tp, Ts.

When the quantities of interest have to be computed in a given spectral range (i.e., over several wavelengths), all the above processes have to be repeated at each single wavelength, according to the way the dispersion is defined. At this point, different practical situations may arise. The simplest case corresponds to having the changes of refractive index with wavelength for one material defined by means of a 'dispersion formula' or 'dispersive model': given the wavelength, we use a formula to calculate the index. Of course, the practical validity of the procedure relies on the previous knowledge of the material. For example, for some dielectric materials in the visible range between 400 and 700 nm (maybe having weak light absorption), a common formula is the Cauchy dispersion [10,11], given by

$$n(\lambda) = n_0 + \frac{n_1}{\lambda^2} + \frac{n_2}{\lambda^4}, \quad k(\lambda) = k_0 \exp\left(\frac{k_1}{\lambda}\right), \quad \lambda \ (nm) \quad (11)$$

(note that $n_0, n_1, n_2, k_0, k_1$ are numerical constants for defining the complex refractive index $N = n - ik$ of the single material which constitutes one single layer, and the notation is not related to the numbers of the layers 0, 1, 2,...). Another common situation arises in case of using one material whose refractive index is known in advance and the corresponding data have been previously stored in a file in form of a list, for our range of interest. We will address these (and other) ways of defining the dispersive refractive index in the Examples.

Finally, there is an important detail to be considered for oblique incidence with dispersive materials regarding the validity of the Snell law in refraction (expression (1)). When we calculate backside quantities, the actual incidence angle on the substrate changes with the wavelength. This fact will be considered explicitly when developing Example 2.

### C. Calculation of the magnitudes of interest

Finally, once the coefficients defined in (6) or (10) are found for the sample, the measurable magnitudes for the problem (usually beam intensity reflectance, transmittance, etc.) may be computed. It is important to note that, for finding the different magnitudes of interest, it is necessary to master the physical concepts involved in the problem (it is not only a numerical issue). For example, assume the incident light beam is linearly polarized, with the electric field vibrating at angle $\phi$ with respect to the incidence plane. Then, the ratio between 'p' and 's' electric field incident amplitudes is $\frac{E_p^0}{E_s^0} = \frac{\cos \phi}{\sin \phi}$. Thus, the reflected and transmitted amplitudes have to be multiplied according to this ratio and the reflected and transmitted energies will be the following weighted means

$$T = T_p \cos^2 \phi + T_s \sin^2 \phi, \quad R = R_p \cos^2 \phi + R_s \sin^2 \phi \quad (12)$$

of course affecting all the above quantities

$$R_p^{a+} \cos^2 \phi, R_p^{a-} \cos^2 \phi, T_p^{a+} \cos^2 \phi, T_p^{a-} \cos^2 \phi,$$
$$R_p^{b+} \cos^2 \phi, T_p^{b+} \cos^2 \phi, R_s^{a+} \sin^2 \phi, R_s^{a-} \sin^2 \phi, \quad (13)$$
$$T_s^{a+} \sin^2 \phi, T_s^{a-} \sin^2 \phi, R_s^{b+} \sin^2 \phi, T_s^{b+} \sin^2 \phi$$

Similarly, for incident elliptically polarized light defined by the parameters $(\psi, \chi)$, it will be necessary to relate them to the 'p' and 's' electric field components and subsequently to the reflected or transmitted calculated magnitudes [9]. Thus, in the most general case, finding the different magnitudes of interest requires solving a detailed physical (not only numerical) problem. The numerical procedures we are explaining in this Tutorial (examples and Appendix) address some common situations. Although is not at all possible to be comprehensive for all the practical configurations and measurable magnitudes, the provided examples illustrate the general guidelines that should be used to tackle for an arbitrary case.

## 3. EXAMPLE 1 (BASIC): SHOWING THE PERFORMANCES OF AN ANTI-REFLECTIVE COATING ON GLASS

### A. Definition of the physical structure

*1. Range of wavelengths*

The starting point of the whole process is to define the range of wavelengths of interest and its units. Subsequently, in all our codes all the calculations will be performed step by step for each single wavelength. The range of wavelengths is fixed by a variable 'x' (here supposed to be in nanometer units), as follows:

```
nlamb=31; x=linspace(400,700,nlamb);
```

The meaning of this line of code is obvious: we take 31 wavelengths equally spaced from 400 to 700 nanometers (both limits included), so in 10 nm steps. It is important to note that the choice of the range and the units is crucial, since this range is implicitly understood to be unique and the same for all the forthcoming procedures. Nevertheless, it should be clear from now that the adaptation of the codes to other units or even to non-equally spaced intervals is straightforward.

## 2. Thickness and dispersion of one layer

Next, we have to define each individual layer of the physical configuration: geometrical thickness and material. We have to decide the units for the thickness. Here we will use 'nm' consistently. To avoid any possible confusion, it is imperative to ensure the same units for the thicknesses and for the wavelengths. To identify each medium (or material) in the lines of code of our scripts we will use two items: i) one keyword acting as the name identifying the material (in practical terms, the type of dispersion of its refractive index), and ii) the numerical parameters of the dispersion type that actually give the refractive index values in the range of interest. For example, for a 120 nm thick layer of material, whose refractive index obeys the Cauchy formula (11) with five parameters $n_0$, $n_1$, $n_2$, $k_0$, $k_1$ with

$$n_0 = 2.0, n_1 = 10000.0 nm^2, n_2 = 0.0, k_0 = 0.01, k_1 = 200.0 nm,$$

we will use a line of code like the following:

```
120. 'Cauchy' {2.0 10000. 0. 0.01 200.};
```

Later in our code, when the material of the layer has been identified by the word `'Cauchy'`, we will use the five parameters supplied, according to the dispersion formula (11) to calculate the refractive index of the material for each wavelength, storing it in a variable 'Nlay'. Namely:

```
v5p(1)=2.0; v5p(2)=10000.; v5p(3)=0.; v5p(4)=0.01;
v5p(5)=200.;
nnn=v5p(1)+v5p(2)./x.^2+v5p(3)./x.^4;
kap=abs(v5p(4))*exp(v5p(5)./x);
Nlay=nnn-1i*kap;
```

Similarly, another common dispersion formula using only two parameters is used for transparent materials [10, 11]. We have shown the first way to specify thickness and refractive index of one layer. Regarding the details of the Matlab language, note the use of the { } braces and blank spacing between values after the text `'Cauchy'` and the use of the operators './' and '.^2' for the division and squaring operations, in order operate for the full length of 'x'. If Python language is used, these conventions are different.

Let us describe the second way to specify thickness and refractive index of one layer. Assume the material for the ambient to be non dispersive. A meaningful line of code for air (for example) can be:

```
Inf 'Constant' {1.000277 0.} ,
```

with 'Inf' meaning infinite thickness and 'Constant' as the appropriate name for designating any material whose refractive index does not change with the wavelength. Here in this line, the ambient could be 'air' with n=1.000277 and kap=0.0. Later the two parameters supplied will be used in the code to store the refractive index values in the variable 'Nlay':

```
v2p(1)=1.000277; v2p(2)=0.;
nnn=v2p(1); kap=abs(v2p(2));
Nlay=(nnn-1i*kap).*linspace(1,1,length(x));
```

This procedure is useful for defining any material without dispersion since it uses two parameters (real and imaginary part of the 'Constant' refractive index) creating an array of constant complex refractive indices, one for each item in 'x'. These indices, in fact, repeat the values supplied 'v2p(1)=1.000277' and 'v2p(2)=0.', but have the right format we use later in our codes. Note that we are forcing the imaginary part of the calculated complex refractive index to be negative, as required in our theoretical developments [2].

Typically we simply use for ambient 'air' the following line:

```
Inf 'Constant' {1. 0.}.
```

Note that for ambient 'water' we could similarly do

```
Inf 'Constant' {1.33 0.}.
```

A third procedure may be the case where the refractive index of the material is known with high precision from the manufacturer. For example, if the coating is deposited on a BK7 glass substrate, the definition of the material as our semi-infinite substrate can be done in a new way, by the code line:

```
Inf 'BK7' {0} ,
```

with the parameter '0' alone meaning that no parameters are required, since we will later use the manufacturer's data [13] for finding the refractive index for the wavelengths ('x') and store them in the variable 'Nlay' as follows:

```
n2=1+(1.03961*x.^2)./(x.^2-6.0e3)+(0.23179*x.^2)./...
  (x.^2-2.0e4)+(1.0146*x.^2)./(x.^2-1.0e8);
nnn=sqrt(n2);
Nlay=nnn-1i*0.0;
```

This code works according to the formula supplied [13]:

$$n^2(\lambda) = 1 + \frac{1.03961\lambda^2}{(\lambda^2 - 6 \times 10^3)} + \frac{0.23179\lambda^2}{(\lambda^2 - 2 \times 10^4)} + \frac{1.0146\lambda^2}{(\lambda^2 - 1 \times 10^8)}, k(\lambda) = 0.0, \lambda \, (nm)$$

Again, we stored in 'Nlay' an array of complex refractive indices, one for each item in 'x', that have null imaginary part.

The fourth way we address corresponds to refractive index data for one material that can be taken from databases or from files storing the values found in previous experiments (for instance, the widely used data for metals from Reference [14]). For our computations, it is only required to have available the values for our wavelength variable 'x'. Thus, when refractive index data are stored in a file, it is convenient to allow defining one single layer with a line of code like

```
135    'File'    {'Ag_n_k.dat'};
```

representing a 135 nm thick layer of a material whose refractive index is contained in a 'File' whose name is 'Ag_n_k.dat', the only requirement being that the data in the file have to cover the range of wavelengths 'x' of our problem. Then, after reading the values, our program will interpolate them linearly to the exact values of the variable 'x' with the Matlab function `'interp1'`. Of course the format of the file to be read has to fulfill some simple requirements.

The interpolating procedure is explained in detail in Appendix B. In Table 1 we summarize the four different ways for introducing the refractive index data of the materials in our codes: 1) the use of a dispersion formula, 2) supplying fixed (constant) values, 3) computing the values from manufacturer's data, 4) use data previously stored in a file.

Table 1. Specifying the dispersion of the materials

| Type | Keyword | Parameters |
|---|---|---|
| Dispersion formula | 'Cauchy' | 5 |
|  | 'Sellmeier' | 2 |
|  | ... | ... |
| Constant | 'Constant' | 2 |
| Manufacturer Data | 'BK7' | 0 |
|  | 'MaterialName' | 0 |
| Data in a file | 'File' | 'Ag_n_k.dat' |
|  |  | 'FileName.dat' |
|  |  | ... |

Additionally, we will present one interesting way for generating a 'File' with refractive index data, valid for mixtures of up to three materials in one single layer. Our procedure uses the Effective Medium Approximation (EMA) [15, 16] for finding the 'effective' refractive index. The code performing these calculations and creating the file (equivalent in format to the previous 'Ag_n_k.dat') is presented in the Appendix C as a script.

*3. The layers of the coating*

Next we have to define our physical configuration (usually a stack of constitutive layers). We plan to study the performances of an anti-reflective (AR) coating on glass. We know from references [2] that three alternating layers of low, high and low refractive indices with suitable thicknesses will do the job. For example, assume we use MgF$_2$ as low index material with dispersion given by the following Cauchy formula with only two coefficients different from zero:

$$n(\lambda) = 1.36 + \frac{4100.}{\lambda^2}, k(\lambda) = 0.0, \lambda \ (nm)$$

and TiO$_2$ as high index material, with dispersion

$$n(\lambda) = 2.0 + \frac{17500}{\lambda^2} + \frac{98000.}{\lambda^4}, k(\lambda) = 0.0, \lambda \ (nm)$$

The complete AR configuration is described by a 'struct' data type in Matlab, that we give the name 'layers', and consists of five lines:

```
layers = {Inf 'Constant'   {1. 0.};
          93  'Cauchy' {1.36 4100. 0. 0. 0.};
          121 'Cauchy' {2.0 17500. 98000. 0. 0.};
          185 'Cauchy' {1.36 4100. 0. 0. 0.};
          Inf 'BK7' {0} };
```

The first layer is in fact the ambient, with thickness 'Inf', material 'Constant' and the 2 parameters for defining the actual refractive index given by the bracket {1. 0.}. The second layer (the most external one) is a 93 nm thick MgF$_2$ layer, whose dispersion is 'Cauchy' type with the five associated parameters {1.36 4100. 0. 0. 0.}. The third (central) layer is a 121 nm thick sheet of TiO$_2$ material again with dispersion type 'Cauchy', but now with the five associated parameters {2.0 17500. 98000. 0. 0.}. The fourth (innermost towards the substrate) layer is like the most external one, but with thickness 185 nm. The 5th layer is, in fact, the semi infinite substrate of BK7 glass material. Note that we have not identified the materials by their names but by their dispersive characteristics, which indeed is the relevant quantity for the practical performances of the setups.

*4. How the dispersion of the layers is handled in the codes*

According to the way we have build our structure 'layers', its line number 1 identifies the ambient, line number 2 designates the most external layer, line 3 the next internal ... increasing towards the substrate. Thus, for 'num_lay'=1, 2, ... the code

```
layers{num_lay,2}
```

corresponds to the keyword (like 'Constant' or 'Cauchy' or 'BK7') that identifies the kind of dispersion for the layer with number 'num_lay'. Thus, once the structure 'layers' and the array of wavelengths 'x' are given, the set of refractive index values 'Nlay' corresponding to the layer with number 'num_lay' is obtained by calling the following function:

```
function Nlay = calc_Nlayer(layers,x,num_lay)
    switch layers{num_lay,2}
        case 'Constant'
            v2p=[layers{num_lay,3}{1},layers{num_lay,3}{2}];
            nnn=v2p(1); kap=abs(v2p(2));
            Nlay=(nnn-1i*kap).*linspace(1,1,length(x));
        case 'Cauchy'
            v5p=[layers{num_lay,3}{1},layers{num_lay,3}{2},layers{num_lay,3}{3},...
                layers{num_lay,3}{4},layers{num_lay,3}{5}];
            nnn=v5p(1)+v5p(2)./x.^2+v5p(3)./x.^4; kap=abs(v5p(4))*exp(v5p(5)./x);
            Nlay=nnn-1i*kap;
        case 'Sellmeier'
            v2p=[layers{num_lay,3}{1},layers{num_lay,3}{2}];
            nnn=sqrt(1+v2p(1).*x.^2./(x.^2-v2p(2))); kap=0;
            Nlay=nnn-1i*kap;
        case 'File'
            aux=load(layers{num_lay,3}{1});  % N,k data
            nnn=interp1(aux(:,1),aux(:,2),x);
            kap=interp1(aux(:,1),aux(:,3),x);
```

```
            Nlay=nnn-1i*abs(kap);
            Nlay=reshape(Nlay,[1,length(x)]);
        case 'BK7'
            n2=1+(1.03961*x.^2)./(x.^2-6.0e3)+(0.23179*x.^2)./ ...
                (x.^2-2.0e4)+(1.0146*x.^2)./(x.^2-1.0e8);
            nnn=sqrt(n2);
            Nlay=nnn-1i*0.0;
    end
end
```

If necessary, adding new types of dispersion formulas or new manufacturers data for materials amounts only to add new 'cases' inside this function, identified by the corresponding keyword (like 'Cauchy', 'Sellmeier' or 'BK7'). The cases 'Constant' and 'File' need the two numerical values or the file name, as explained above.

### B. Performing the core calculations

We will assume always the ambient to be transparent [2] but, to simplify computations, we use one incidence angle and one ambient refractive index value for each single wavelength in the array 'x'. Thus, both the index of the transparent ambient ('N0') and the incidence angle in radians ('incang') must be arrays of length 'nlamb'.

According to the theory presented in Section 2., for one layer of thickness 'd' whose dispersive refractive indices are 'N', the computations use the 2x2 matrices mentioned in our expression (3). They are different for each wavelength in 'x' and are named Ms for 's' and Mp for 'p' polarizations. The matrix product (for each wavelength in 'x') for the whole structure 'layers' is performed and the overall 'rs' and 'rp' (and also Ts and Tp) coefficients of the assembly are obtained by means of the following function:

```
function[rs,rp,Ts,Tp]=calc_rsrpTsTp(incang,layers,x)
```

Most of the measurable quantities such as reflectance, transmittance, polarization changes... are directly computed from the knowledge of these overall [rs,rp,Ts,Tp] coefficients. The source code and a brief description of this function is presented in the Supplementary Material. In fact this function contains the core of all the calculations. The user has to set the incidence angle, the structure of the layers and the wavelengths (the arguments of the function) before calling it. But the good thing is that the user never has to modify the 'calc_rsrpTsTp' function.

### C. Plotting the spectral reflectances

Example 1 will illustrate the calculations of light reflectances for one stack of three layers on a semi-infinite substrate. Semi-infinite is the term for a substrate where there is no beam coming back from the rear of the substrate. According to our expression (7), the intensity reflectances Rs or Rp can be computed by simply squaring the values 'rs' and 'rp' previously found.

The complete code for the example is in the Supplementary material. First we define the wavelength range, the physical configuration and the incidence angle. Then we will calculate the reflectance at normal incidence, i.e., for the array of angles

```
incang=0. * linspace(1,1,length(x)),
```

where 'p' and 's' polarizations are equivalent [2]. For comparison, we will show the 'Rp' and 'Rs' reflectances at 40 deg, i.e., by making

```
incang=40.*pi/180 * linspace(1,1,length(x)).
```

Moreover, the reflectance of the bare substrate at normal incidence is also calculated by redefining the configuration (i.e., eliminating the coating).

Figure 3 illustrates how the reflectance of the bare substrate at normal incidence (Rglass), that is more than 4% for all the full range from 400 to 700 nm, is reduced to a mean value of around 1% for that range (line R0). For comparison, for the incidence angle of 40 deg, the two quite different reflectances Rs40 and Rp40 are also shown.

Next in this Example 1, for the 40 deg. incidence angle, we compare the light reflection for the two following cases: incident linearly polarized light (vibrating at 30 deg. with respect to the incidence plane) and incident natural (non-polarized) light. Figure 4 shows these results, designated respectively as 'R40P30' and 'R40Nat'. They have been computed using our expression (12).

It is important to note that the quantities 'Ts' and 'Tp' computed here correspond to the light that has entered into the substrate and, thus, are not directly measurable. At normal incidence the transmittance through the two sides of a plano-parallel substrate of refractive index 'n' in air can be approximately estimated as the product of our 'Ts' (or 'Tp' since incidence is normal) by the transmittance of one substrate-ambient interface (given by the quantity $4n/(1+n)^2$ ).

In Example 2 we cover in detail how to measure on double sided substrates.

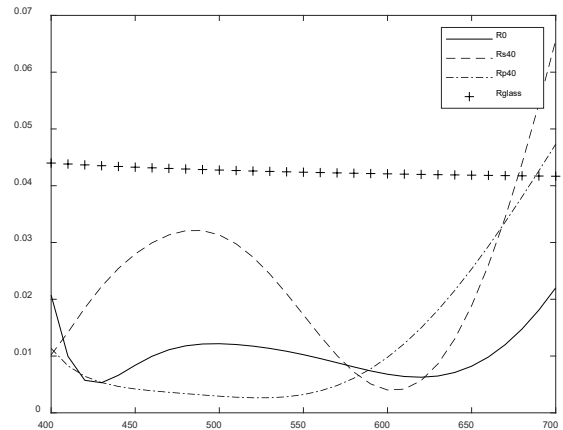

Figure 3. Reflectances of a semi infinite glass substrate coated with our three thin layers. Reflectance of the sample (coated glass) 'R0' and of the bare substrate 'Rglass' at normal incidence. For 40 deg. incidence angle, reflectances of the sample for incident 's' polarisation 'Rs40' and for incident 'p' polarization 'Rp40'.

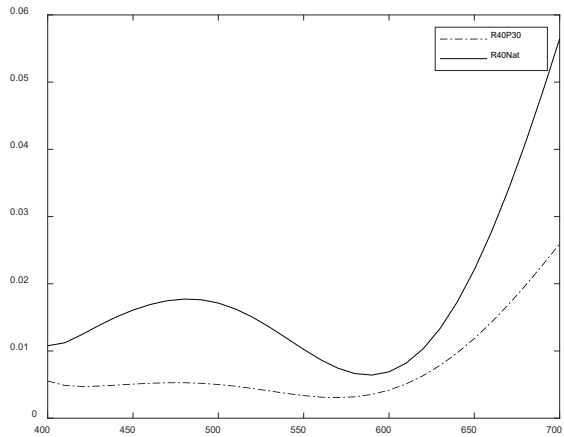

Figure 4. For 40 deg. incidence angle, reflectances of the sample for incident linearly polarized light (vibrating at 30 deg with respect to the incidence plane) 'R40P30' and for incident non polarized (natural) light 'R40Nat'.

## 4. EXAMPLE 2 (ADVANCED): COMPUTATIONS FOR A SUBSTRATE COATED AT BOTH SIDES

### A. Definition of the physical structure

*1. The coatings of the two sides*
For Example 2, assume the substrate is a plano parallel slab of transparent glass, coated with thin films at both sides. The transparency of the substrate is an absolute requirement for the validity of the following two-sided calculations. Certainly, any substrate with a thickness of the order of millimeters will absorb all the light unless the imaginary part of its complex refractive index value (written 'k' in our formulas above) is absolutely negligible. Thus, the results computed in our codes are not valid unless the requirement of transparency of the substrate is fulfilled.
For simplicity, we continue working with the same range of wavelengths:

```
nlamb=31; x=linspace(400,700,nlamb);
```

If the substrate is transparent, its thickness is not relevant and it can be characterized in our codes as one layer of 'Inf' thickness:
```
Inf 'BK7' {0};
```
Assume that the substrate is coated at the first side, and the first side structure is (from ambient to substrate):

```
layers_a = {Inf 'Constant' {1. 0.};
            200 'Sellmeier' {1.7 10000.};
            30  'Cauchy'    {1.5 10000. 0. 0.1 150.};
            Inf 'BK7'       {0} };
```

We will suppose that the second side is also coated, with a structure (from substrate to ambient):

```
layers_b = {Inf 'BK7'      {0};
            300 'Cauchy'   {1.8 10000. 0. 0. 0.};
            30  'File'     {'EMA3_n_k.dat'};
            Inf 'Constant' {1. 0.} };
```

Note that, for the problem to be meaningful, the last layer at the incident side must be exactly like the first layer of the second side.

### B. Performing the core calculations

*1. Front side ('a') forward calculations*
Assuming an incidence angle of 10 degrees, the following code calculates the 'Plus' quantities at the 'a' side, using expression (10):

```
incang=10*pi/180*linspace(1,1,length(x));
[rs,rp,Ts,Tp]=calc_rsrpTsTp(incang,layers_a,x);
Rs_aPlus=(abs(rs)).^2;
Rp_aPlus=(abs(rp)).^2;
Ts_aPlus=Ts;
Tp_aPlus=Tp;
```

*2. Front side ('a') backward calculations*
To calculate the 'Minus' quantities at the 'a' side at oblique incidence is a subtle matter. First, reversing the layers is required, by defining a new configuration:

```
layers_ainv=layers_a(end:-1:1,:);   % inverts layers
```

where the use of the ':' operator does the reversing inside the parenthesis '(end:-1:1,:)'.

A deep understanding of the electromagnetic theory explained in chapter 2 of Ref [2] would be required for fully demonstrating the following procedure. In practical terms, it is only necessary to note that the dispersive nature of the materials implies that the angle entering the substrate changes with the wavelength, according to Snell law (1). Thus, we need to calculate the incidence angle in the way back ('incang_inv'), from the knowledge of the forward incidence angle ('incang'). The refractive indices of the 'ambient' medium in the 'forward' and 'backward' trajectories, respectively, are:

```
N0=calc_Nlayer(layers_a,x,1);
N0_inv=calc_Nlayer(layers_ainv,x,1);
```

If Snell law (1) is applied to the first and last angles at the coating of the 'a' side, one has:
$$N_0 \sin \vartheta_0 = N_m \sin \vartheta_m,$$
where the quantities $N_0$, $\vartheta_0$, $N_m$ and $\vartheta_m$ correspond in our code, respectively, to the variables 'N0', 'incang', 'N0_inv' and 'incang_inv'. Thus the code for computing the right value for 'incang_inv' is:

```
incang_inv=asin(N0.*sin(incang)./N0_inv);
```

Now we are ready for the backwards (corresponding to the 'Minus') calculations at the 'a' side

```
[rs,rp,Ts,Tp]=calc_rsrpTsTp(incang_inv,layers_ainv,x);
Rs_aMinus=(abs(rs)).^2;
Rp_aMinus=(abs(rp)).^2;
Ts_aMinus=Ts;
Tp_aMinus=Tp;
```

### 3. Back side ('b') calculations

Fortunately for the 'b' side we only need the forward calculations. Since the substrate is plano-parallel, the incidence angles for the wavelengths 'x' at side 'b' in the forward direction are the same as the exit angles at side 'a' in forward direction. These are exactly the same angles we have just calculated: `incang_inv`. Thus, for the forward direction at side 'b', where the coating is defined by the structure `layers_b` the required code is

```
[rs,rp,Ts,Tp]=calc_rsrpTsTp(incang_inv,layers_b,x);
Rs_bPlus=(abs(rs)).^2;
Rp_bPlus=(abs(rp)).^2;
Ts_bPlus=Ts;
Tp_bPlus=Tp;
```

## C. Combining and plotting the results

### 1. Computing the overall R and T

All the required quantities are available now for computing the measurable magnitudes R or T for the substrate coated at both sides. Anyway, it has to be noted that the 's' and 'p' modes are still independent and separate calculations have to be performed, according to expressions (9-10) as follows:

```
Rs=(Rs_aPlus+Rs_bPlus.*(Ts_aPlus.^2-Rs_aMinus.*Rs_bPlus)) ./ (1-Rs_aMinus.*Rs_bPlus);
Rp=(Rp_aPlus+Rp_bPlus.*(Tp_aPlus.^2-Rp_aMinus.*Rp_bPlus)) ./ (1-Rp_aMinus.*Rp_bPlus);
Ts=(Ts_aPlus.*Ts_bPlus) ./ (1-Rs_aMinus.*Rs_bPlus);
Tp=(Tp_aPlus.*Tp_bPlus) ./ (1-Rp_aMinus.*Rp_bPlus);
```

### 2. Some interesting comparisons

One can now compare the results. It is interesting for example to illustrate the difference between the reflectance of the sample (coated at both sides) with the reflectance due to the first side alone. Note that this last quantity can be easily measured by grinding the back side of the substrate (in fact destroying the glass smoothness and the coating at the back side), to avoid the light reflection from this side. Note that the differences between 's' and 'p' polarizations in Figure 5 are indeed small, as expected due to the small value of the angle of incidence (10 deg). Conversely, the same Figure 5 illustrates the big differences between the reflectances due to the 'a' side alone (Rsa and Rpa) and the reflectances due to the effect of the 'a' and 'b' sides together (Rs and Rp).

It is also interesting to check the validity of a completely general result demonstrated in the electromagnetic theory of thin films [2]: any 'Plus' transmittance coincides with the corresponding 'Minus' one. This checking is done in Figure 6, for the 's' and 'p' independent polarisations. The following code corresponds to to the two Figures just mentioned

```
figure, plot(x,Rs,'k.-',x,Rs_aPlus,'k:',x,Rp,'k--',x,Rp_aPlus,'k-.'), legend('Rs','Rsa','Rp','Rpa')
figure, plot(x,Ts_aPlus,'k.',x,Ts_aMinus,'k:',x,Tp_aPlus,'k--',x,Tp_aMinus,'k-.'),
        legend('TsaP','TsaM','TpaP','TpaM')
```

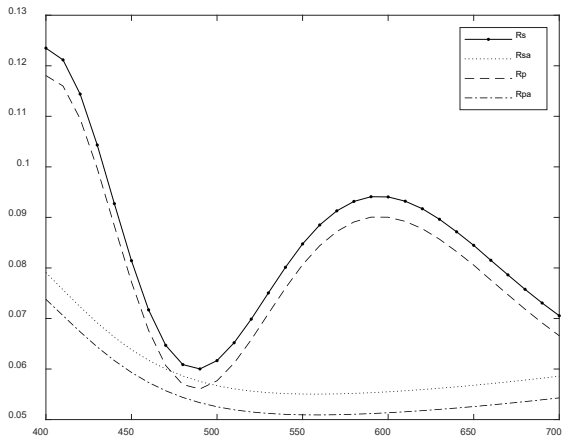

Figure 5. For 10 deg. incidence angle, reflectances due to the two sides of the coated substrate: 'Rs' for incident 's' polarized light, 'Rp' for 'p' polarized light. For 10 deg. incidence angle, reflectances due to the 'a' side alone of the coated substrate: 'Rsa' for incident 's' polarized light, 'Rpa' for 'p' polarized light.

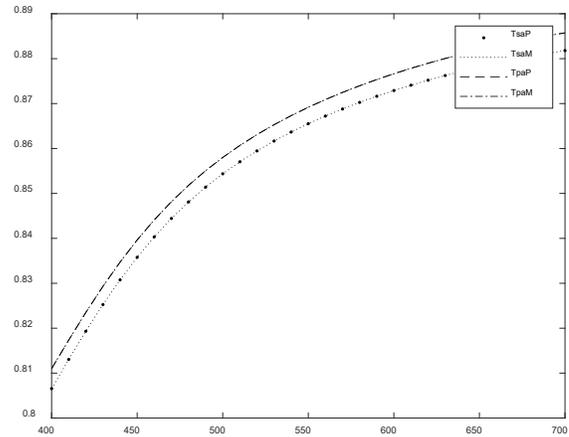

Figure 6. Computed transmittances for the 'a' side of the substrate, whose coating includes an absorbing layer. One can see the differences between 's' and 'p' polarisations and the equivalence between the 'Plus' and 'Minus' propagation directions, since 'TsaP' is the same as 'TsaM' and 'TpaP' is the same as 'TpaM'.

Figure 6 illustrates that, even when our coating include absorbing layers (like the innermost 'Cauchy' type one at side 'a'), our 'Plus' transmittance values coincide with the corresponding 'Minus' transmittances. Thus, 'TsaP' coincides with 'TsaM' and 'TpaP' coincides with TpaM. Note that this is not the case for reflectances.

## 5. EXAMPLE 3 (RESEARCH): SURFACE PLASMON POLARITON

### A. Definition of the physical structure

Surface plasmons are surface charge oscillations propagating along a metal-dielectric interface when excited by proper illumination conditions [17–19]. Surface plasmons are widely exploited for sensing purposes because the charge oscillation is highly dependent on the metal-dielectric interface shape, excitation configuration and optical properties of metal and dielectric. One of the most common ways to excite surface plasmons is the so-called Kretschmann configuration, in which a metal layer is deposited on top of a (hemi-spherical) glass substrate. When the film is illuminated from the glass side, a surface plasmon can be excited at the metal-air interface when light is incident with the proper angle. Assuming monochromatic light excitation (633. nm), we will simulate this configuration considering the propagation of light on the structure:

```
layers =  {Inf 'Constant' {1.5 0.};
            30 'Constant' {0.056206 4.2776};
           Inf 'Constant' {1. 0.}};
```

where we consider a 30 nm thick silver layer on top of the glass. Note that, although the glass is the physical substrate, the layer structure is built according to the direction of light propagation (glass/layer/air). Since we are dealing with monochromatic light, we use the model 'Constant' for all involved materials. First we compute the reflectance of the system for any incidence angle

```
x = 633.0;
numang = 1000;
Rs=zeros(1,numang); Rp=zeros(1,numang);
angles = linspace(0,90,numang)*pi/180;
cont=0;
for incang = angles
   cont=cont+1;
   [rs,rp,Ts,Tp]=calc_rsrpTsTp(incang,layers,x);
   Rs(cont)=abs(rs)^2; Rp(cont)=abs(rp)^2;
end
figure, plot(angles*180/pi,Rs,angles*180/pi,Rp);
xlabel('incidence angle (deg)'),ylabel('Reflectance'),
legend('Rs','Rp')
```

Note that, contrary to the previous examples, the wavelength vector array entering in the calculation has a constant value while the angle of incidence does not. We can now plot the dependence of reflectance with the angle of incidence.

The figure shows that the reflectivity values are quite high (as corresponding to reflection from a metal surface), except for the case of p-polarized light incising at 43.6°, where a narrow drop in reflectance is observed. It corresponds to the resonant excitation of a surface plasmon at the metal-air interface: the incident light energy is efficiently transferred to the surface charge oscillation instead of being reflected to the detector. Note that the excitation is only possible for p-polarized light, for surface plasmons are longitudinal waves and couple with the electric field component perpendicular to the interface.

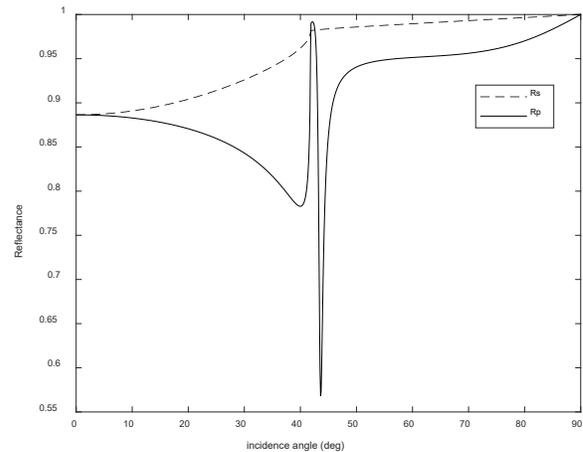

Figure 7. Reflectances of a glass/silver/air system as a function of the incidence angle.

### B. Refining the configuration for best analysis

It can intuitively be expected that if the metal layer is too thick, light will not reach the metal-air interface and no surface plasmon will be excited. On the other hand, a too thin metal layer may be unable to produce efficient surface charge oscillations and an optimal metal layer thickness for surface plasmon excitation should exist. We can proceed as follows to find out the optimal thickness. We start by defining the same configuration, but now we will reduce the range of incidence angles and will change the metal layer thickness values, say between 0 and 100 nm. For these two ranges of values we will calculate Rp (the most relevant quantity, as explained) by using a loop over all metal layer thickness values: at each step we change the value of the thickness by modifying the quantity 'layer{2,1}'. The results are plotted using the 'pcolor' function.

```
% x = 633.0;
numang = 150; angles = linspace(35,50,numang)*pi/180;
numthk = 100; thicks = linspace(0,100,numthk);
Rp=zeros(numthk,numang);
cont=0;
for incang = angles
   cont=cont+1;
   for ithk = 1:numthk
      layers{2,1}=thicks(ithk);
      [rs,rp,Ts,Tp]=calc_rsrpTsTp(incang,layers,x);
      Rp(ithk,cont)=abs(rp)^2;
   end
end
figure, pcolor(angles*180/pi,thicks,Rp), colormap gray
xlabel('incidence angle(deg)'), ylabel('thickness'),
colorbar, shading interp
```

We see that the deepest drop in reflectance is observed for a layer thickness around 50 nm. As expected, very thick layers give high reflectivity and very thin layers show the total internal reflection behaviour of a glass/air interface with the characteristic critical angle around 41.8°.

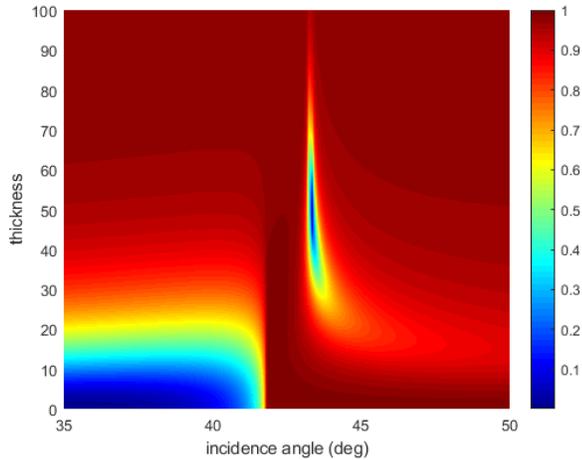

Figure 8. Reflectances of a glass/silver/air system versus silver thickness and incidence angle.

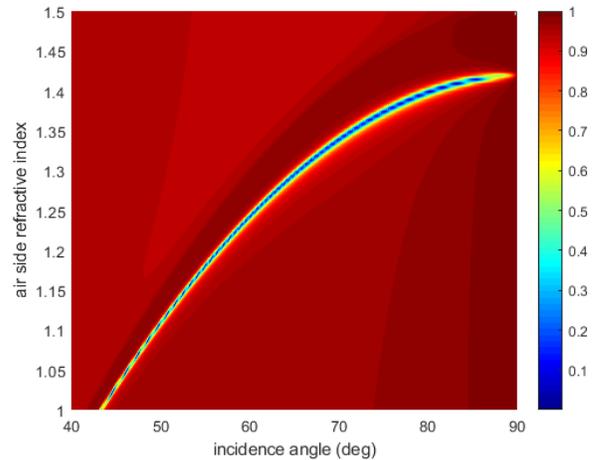

Figure 9. Reflectances of our sample versus air side refractive index and incidence angle.

Finally, we can evaluate how the surface plasmon resonance in the Kretschmann configuration can be used for sensing of analytes present on the air side of the metal layer. Thus, we will analyze how the reflectance angular depencence changes when the refractrive index of the air side varies due to the analyte presence. We will proceed in the same way as before, but instead of varying the thickness (that will be fixed now to 50 nm) we will vary the refractive index of the last layer ("substrate") of the layers structure.

```
% 50 nm
layers =  {Inf 'Constant' {1.5 0.};
           50 'Constant' {0.056206 4.2776};
           Inf 'Constant' {1 0};};

numang = 200; angles = linspace(40,90,numang)*pi/180;
numnout = 100; nout = linspace(1.0,1.5,numnout);
Rp = zeros(numnout,numang);
cont=0;
for incang = angles
    cont=cont+1;
    for inou=1:numnout
        layers{3,3} = {nout(inou),0};
        [rs,rp,Ts,Tp]=calc_rsrpTsTp(incang,layers,x);
        Rp(inou,cont) = abs(rp)^2;
    end
end
figure, pcolor(angles*180/pi,nout,Rp), colormap gray
xlabel('incidence angle (deg)')
ylabel('air side refractive index')
colorbar, shading interp;
```

We can see that, as the refractive index of the air-side changes, the position of the reflectance dip shifts to larger angles. For small refractive index changes, the variation is nearly linear and we can define a sensitivity factor of 60 degrees per refractive index unit. In actual sensing experiments it could be more realistic to consider a water-like environment, with refractive index variations in a small range around the refractive index of water (1.33). Using these scripts we could easily compare the performance of different materials or even material mixtures.

## 6. FINAL SUMMARY OF THE PROCEDURES

### A. Preliminaries

The present Tutorial develops an intuitive approach to the topic of numerical computations for thin film coatings using a high level programming language environment (like Matlab or Python). Thus, after a concise theoretical introduction, we guide the user through practical examples. We want to enable the user to adapt the codes presented here to his/her needs. Let us summarize the full procedure.

Before starting the process for the computations, it is necessary to know in detail the physical configuration of the sample: incidence angle and complete physical configuration, including physical thickness of all the layers and also all the refractive indices involved in the range of wavelengths of interest (including the materials for the ambient and the substrate).

The refractive indices of the materials can be defined in several ways and must be identified with a keyword, according to the dispersive characteristics of the material (see Table 1). The ambient must be transparent. Usually its dispersion type will be 'Constant' and we have to supply the two values $n_0$ and $k_0 \equiv 0.0$ for its complex refractive index $N_0 = n_0 - ik_0$. In fact, the ambient could be dispersive (provided all $k_0 \equiv 0.0$). Table 1 summarizes the different ways for defining the dispersion, and the details have been explained for Example 1.

Thus, before starting the computations, we must be sure that the function 'calc_Nlayer' explained above, includes all the 'cases' that correspond to all the materials present in our configuration. If necessary, it is straightforward to include more 'cases' as required. Implicitly, we also must be sure that the files to be called by the function are appropriately defined and are available at the folder where we will work. Note that the function 'calc_Nlayer' can be called directly by the user within the script to, for instance, plot the complex refractive index dispersion over the spectral range of interest.

Once the above function is properly defined, we may be sure that the later calculation of the coefficients [rs,rp,Ts,Tp] will work.

In effect, the function `'calc_rsrpTsTp'` first, will calculate the refractive indices for the materials in the structure `'layers'` by calling the above function `'calc_Nlayer'`. Subsequently, this same function `'calc_rsrpTsTp'` will perform the calculations corresponding to our expressions (3), (4) and (5) automatically.

### B. Creating the script for the calculations

Finally, the required script for the actual calculations, as in our detailed examples, will have to cover the following stages, in order:
- define the range of wavelengths.
- define the physical structure of the sample.
- call the function `'calc_rsrpTsTp'`.
- perform the final calculations, always derived from knowing `'rs'`,`'rp'`,`'Ts'` and `'Tp'`, as illustrated in the examples or in Appendix A.

## 7. EXPANSION OF THE TOPICS COVERED. FUTURE EXTENSIONS

There are many practical situations not addressed in the present tutorial, mainly because they would require sets of additional specifications. For example, dealing with the case of slightly absorbing substrates or coatings with inhomogeneous layers [20, 21].

The major general limitation of our work is the following: in the present Tutorial we have always addressed only the 'direct problem': given all the quantities that uniquely define the physical problem, we have explained how to calculate the corresponding measurable magnitudes. But there is another kind of advanced software procedures that we have not addressed at all: the 'inverse problem' where the unknowns are some values that define the configuration itself. These unknowns have to be found (estimated) from the fitting of experimental data that are available to solve the problem.

A practical case where a basic inverse problem appears is, for example, the manufacture of an AR coating prototype. Starting from the design values for the thicknesses of the layers, a vacuum chamber is prepared for coating the substrates. Modern equipment requires only introducing on a keyboard the parameters for the thicknesses of the layers, but the whole process implies some calibration, until the final results for the actual thicknesses are satisfactory. This final 'quality' is assessed by measuring the reflectance obtained, and then changing the parameters for the thicknesses by trial-and-error procedure, until the performance is satisfactory. The estimation of the thicknesses during the process is done from the reflectance measurements, solving a typical 'inverse problem'. The way to address inverse problems in software packages is to define a 'merit function' and to use a suitable search algorithm that wisely changes iteratively the unknown quantities until a minimum for the merit function is found [22].

We intentionally have not addressed the procedures required for solving inverse problems since they would be too specific for an introductory Tutorial. Moreover, it would be very difficult to propose general techniques for the wide range of potential optimization objectives. There are specialized software packages that address the computation and fitting of performances (including design features) for a wide range of thin film structures [23–25]. Anyway, the ability for addressing simple (not highly specialized) numerical design problems is indeed of great practical interest in laboratories where the manufacture of thin film products is the daily task. Often it is not useful to rely on very specialized software packages, mainly when sophisticated new designs are not required. In conclusion, ordinary laboratory tasks involving simple 'inverse problems' could be the subject of another Tutorial, covering topics like the characterization of the materials in thin film phase, the development of simple optical design techniques for quick manufacture, the optimization of the geometry for improving thickness uniformity, etc [20,26–28].

## APPENDIX A: OTHER CALCULATIONS OF PRACTICAL INTEREST

On the basis of the same numerical scheme we have developed in this Tutorial, many other calculation of practical interest can easily be made. For example, for a thin film coating on a semi-infinite substrate, it may be of interest to find the change of phase between the incident and the reflected beams. We have not covered this kind of topics here but it is not difficult to show [9, 29, 30] that we can address them in terms of the magnitude $\rho = r_p/r_s$, or $(\psi, \Delta)$, defined in our expression (8). The relevant values can be directly calculated from our functions. Namely,

```
RO = rp./rs;
cosDelta = -cos(angle(RO));
tanPsi = abs(RO);
```

As a second example, if the substrate is not a perfectly parallel slab of material, when measuring the light transmittance it may be more accurate to neglect multiple reflections at the sides of the substrate (an implicit assumption for the validity of expression (9)) and make the hypothesis that the actual measurement is not given by
`T = (T_aPlus.*T_bPlus) ./ (1-R_aMinus.*R_bPlus);`
which is expression (9), but more correctly by, simply,
`T = T_aPlus*T_bPlus;`
Clearly, in this second example we illustrate a practical problem where the theoretical formulas are different and the key point for the success of the laboratory procedures is to know what is the right model to represent our actual sample.

As a final example, sometimes the substrate is a piece of (expensive) polished material not to be spoiled by grinding its back surface, as required for measuring the one side reflectance. Then, it may be useful to measure the reflectance from the back side. Strictly speaking, the procedures developed for Example 2 do not include the option we need. But, to calculate the back side reflectance from the knowledge of the front side configuration may be done as follows. Suppose we have the following configuration

```
layers_a = {Inf 'Constant' {1. 0.};
            200 'Sellmeier' {1.7 10000.};
            100 'Cauchy'    {1.5 10000. 0. 0. 0.};
            Inf 'BK7'       {0};

layers_b = {Inf 'BK7'    {0};
            Inf 'Constant' {1. 0.}};
```

Then, by changing the configuration to the following

```
layers_a = {Inf 'Constant' {1. 0.};
            Inf 'BK7'      {0}};

layers_b = {Inf 'BK7'      {0};
            100 'Cauchy'   {1.5 10000. 0. 0. 0.};
            200 'Sellmeier' {1.7 10000.};
            Inf 'Constant' {1. 0.}};
```

the use of expressions (9) solves the problem.

## APPENDIX B: INTERPOLATION OF REFRACTIVE INDEX DATA

Suppose that the refractive index of one material is to be obtained from interpolation of values stored in a database of optical constants for coating materials [14, 31, 32]. In this case, the simplest way for using our software is linear interpolation of the refractive index values (for the current range of wavelengths). The file with the data must have the suitable format. For example, assume we have the original data for 'Ag' in a file named `'Ag_n_k.dat'`, that covers the 188-1937 nm wavelength range (find it with our supplementary material). The file contains refractive index data for 'Ag' in the 3 columns, with the following format ('nm  n  k'), where the 'k' values are positive:

...
276.1          1.41          1.331
284.4          1.41          1.264
...

Provided the data cover the wavelength range, the wavelength units are the same as in our code (in our case 'nm') and they are stored in three columns as shown, our code reads the file (here `'Ag_n_k.dat'`) and linearly interpolates the original 'n' and 'k' values to our wavelengths of interest, as defined by the variable 'x'. The suitable lines of code correspond to the "`case 'File'`" contained inside the function `'calc_Nlayer'` of the text above.

## APPENDIX C: CHANGING THE RANGE OF WAVELENGTHS AND UNITS

The user has to take care of the ranges covered by data when interpolation procedures are involved or when units are mixed. For simplicity, the range of wavelengths and the units for them have not changed in our Examples 1 and 2. Anyway, it should be clear that these kind of variations may be easily implemented. The key point is to realize that the number of elements and the interval between them in the variable 'x' are not relevant for all the subsequent procedures, since 'x' is an input argument for the functions. Thus, for example, we have made

```
nlamb=31; x=linspace(400,700,nlamb);
```

but, replacing this line by the following

```
x=[420., 470., 540., 590., 600., 670.];
nlamb=length(x);
```

all our codes would equally work for these six scattered values. Changing the scripts and functions for other thickness units is conceptually simple, but may be the origin of important practical problems, since it would require modify the functions at the points where the thickness 'd' is multiplied or divided by the wavelengths in 'x'. Thus, for simplicity, we strongly recommend to perform the conversion of units at the very beginning and at the end if needed.

## APPENDIX D: REFRACTIVE INDEX FOR MIXTURES OF MATERIALS. THE EMA THEORY

The Effective Medium Approximation (EMA) [15, 33] is an electromagnetic theory for representing the refractive index of mixtures of materials. When only two materials are combined, the resulting complex refractive index can be analytically calculated from the values and proportions of the components. When more than two materials are mixed, a numerical approach is required for calculations. The script included in the supplementary material illustrates how to use the EMA theory for mixtures up to three materials, applying the method explained in the Ref [16]. It works as follows. For an array of wavelengths defined in the usual form, here named 'x', the script creates a file `'File_EMA_3.dat'` containing the refractive index data computed by an EMA model from the porous mixture of two materials (Ag and $SiO_2$) whose refractive indices are stored in the two files `'Ag_n_k.dat'` and `'SiO2_n.dat'`. Since the mixture is porous, we will model it as made by the two materials plus a third one that is 'Vacuum' (with refractive index equal to 1). The volume fractions of the three constitutive materials are named in the script, respectively, `fr(1)`, `fr(2)` and `fr(3)`. For better illustration of how to read files and on how to manage minor reading variations in the script, the input data are effectively read differently for the two materials. For 'Ag' the data are read from a 'nm-n-k' three-column file format. We always assume that the 'k' values we read and write in files are positive. To be consistent with our theoretical approach, when the call the function '`calc_Nlayer`' is done inside the function '`calc_rsrpTsTp`'`, the values 'Nlay' have negative imaginary part as required. For 'SiO2' only the 'nm-n' data are read from the file, because it is known in advance that all 'k' values are '0'. For the 'Vacuum', the index is in fact not read and simply is assumed to be a constant value '1.0'. The 'Ag' data are contained in a file named `'Ag_n_k.dat'`. Similarly, a file named `'SiO2_n.dat'` for 'SiO2' data can be found in our supplementary material. Of course, the user has always to take care about the wavelength units and has to check that the intervals of the data read is enough to cover the range of 'x' in our codes. Here, for 'Ag' the data cover the 0.18-1.9 micron range and for 'SiO2' cover the 0.28-0.83 micron range. In the example below we define 'x' from 320 to 790 nm. In Examples 1 and 2, we need only to have the range 400-700 nm covered. The example of part 6. of Supplementary material finally writes on the active folder of the computer a dispersive 'File' material, ready to be used in all our codes, according to the procedure introduced in Example 2, section 4.A. Note that the present script has to include at the bottom of the Matlab file the function '`function sumabs = fmerit_EMA(V)`', that, for each wavelength, computes the effective complex refractive index as the best fit (given by the EMA theory for the constituent refractive indices) when used together with the Matlab function 'fminsearch' in the way shown here.

**Acknowledgments.** The authors wish to thank Drs. Enrico Massetti, Angela Piegari, Detlev Ristau, Stefan Guenster, Adolf Canillas and all the colleagues in projects CHRX-CT93-0336 and FMRX-CT97-0101 for their friendship and former collaboration. Prof. Angus H. Macleod trespassed during the writing of this Tutorial. Requiescat in pace.

# Thin film optics computations in a high-level programming language environment: supplement

## SALVADOR BOSCH,[1,*] JOSEP FERRÉ-BORRULL,[2] JORDI SANCHO-PARRAMON[3]

[1]*Universitat de Barcelona, Dep. Física Aplicada, Martí i Franqués 1, 08028 Barcelona*
[2]*Universitat Rovira I Virgili, Eng. Electrònica, Elèctrica i Automàtica, Campus Sescelades, Avinguda Països Catalans, 26, 43007 Tarragona*
[3]*Ruder Bosckovic Institute, Laboratory for optics and optical thin films, Bijenicka cesta 54, 10000 Zagreb*
*\*Corresponding author: sbosch@ub.edu*

---

This supplement gives complete details and listings of the code of the 'functions' and the data files mentioned and used in the text of the Tutorial.

**1. EXPLANATIONS FOR function Nlay=calc_Nlayer(layers,x,numlay)**

Given the structure 'layers' and the array of wavelengths 'x', the set of refractive index values 'Nlay' corresponding to the layer with number 'num_lay' is obtained by calling the following function, where the materials are identified by their dispersive characteristics by means of a keyword, as explained in the main text:

```
% calc_Nlayer.m
function Nlay = calc_Nlayer(layers,x,num_lay)
    switch layers{num_lay,2}
        case 'Constant'
            v2p=[layers{num_lay,3}{1},layers{num_lay,3}{2}];
            nnn=v2p(1); kap=abs(v2p(2));
            Nlay=(nnn-1i*kap).*linspace(1,1,length(x));
        case 'Cauchy'
            v5p=[layers{num_lay,3}{1},layers{num_lay,3}{2},layers{num_lay,3}{3},...
                layers{num_lay,3}{4},layers{num_lay,3}{5}];
            nnn=v5p(1)+v5p(2)./x.^2+v5p(3)./x.^4; kap=abs(v5p(4))*exp(v5p(5)./x);
            Nlay=nnn-1i*kap;
        case 'Sellmeier'
            v2p=[layers{num_lay,3}{1},layers{num_lay,3}{2}];
            nnn=sqrt(1+v2p(1).*x.^2./(x.^2-v2p(2))); kap=0;
            Nlay=nnn-1i*kap;
        case 'File'
            aux=load(layers{num_lay,3}{1});  % N,k data
            nnn=interp1(aux(:,1),aux(:,2),x);
            kap=interp1(aux(:,1),aux(:,3),x);
            Nlay=nnn-1i*abs(kap);
            Nlay=reshape(Nlay,[1,length(x)]);
        case 'BK7'
            n2=1+(1.03961*x.^2)./(x.^2-6.0e3)+(0.23179*x.^2)./ ...
                (x.^2-2.0e4)+(1.0146*x.^2)./(x.^2-1.0e8);
            nnn=sqrt(n2);
            Nlay=nnn-1i*0.0;
    end
end
```

Adding new types of dispersion formulas or new manufacturers' data for materials amounts only to add new 'cases' inside this function, identifiyng them by some new keyword.

## 2. EXPLANATIONS FOR function [rs,rp,Ts,Tp]=calc_rsrpTsTp(incang,layers,x)

For an incidence angle array 'incang', given an array of wavelengths 'x' and a physical configuration 'layers', defined in the way explained, the function computes the quantities [rs,rp,Ts,Tp], corresponding to the magnitudes explained in the text (in our notation), implementing the formulas 2.113 and 2.115 of Reference [2] in the main text. The output of the function is in the form of four arrays of complex values, each one of length 'nlamb'. For a practical use of our codes it is not required to modify any line in the function (it is only presented here for completeness).

```
% calc_rsrpTsTp.m

function [rs,rp,Ts,Tp] = calc_rsrpTsTp(incang,layers,x)
    Ms=eye(2); Mp=eye(2);
    for ix=1:length(x)
        Ms(:,:,ix)=eye(2); Mp(:,:,ix)=eye(2);
    end
    im=1;
    N0=calc_Nlayer(layers,x,im);
    N0s=N0.*cos(incang); N0p=N0./cos(incang);
    for im=2:size(layers,1)-1
        Nlay=calc_Nlayer(layers,x,im);
        ARR=sqrt(Nlay.^2-N0.^2.*(sin(incang).^2));
        Ns=complex(abs(real(ARR)),-abs(imag(ARR)));
        d=layers{im,1};
        Dr=2*pi.*d./x.*Ns;
        Np=Nlay.^2./Ns;
        for ix=1:length(x)
            S(:,:,ix)=[cos(Dr(ix)) 1i./Ns(ix).*sin(Dr(ix));...
                1i.*Ns(ix).*sin(Dr(ix)) cos(Dr(ix))];
            P(:,:,ix)=[cos(Dr(ix)) 1i./Np(ix).*sin(Dr(ix));...
                1i.*Np(ix).*sin(Dr(ix)) cos(Dr(ix))];
            Ms(:,:,ix)=Ms(:,:,ix)*S(:,:,ix); Mp(:,:,ix)=Mp(:,:,ix)*P(:,:,ix);
        end
    end
    im=size(layers,1);
    Nm=calc_Nlayer(layers,x,im);
    ARR=sqrt(Nm.^2-N0.^2.*(sin(incang).^2));
    Nms=complex(abs(real(ARR)),-abs(imag(ARR)));
    Nmp=Nm.^2./Nms;
    for ix=1:length(x)
        V_s = Ms(:,:,ix)*[1; Nms(ix)]; Bs=V_s(1); Cs=V_s(2);
        rs(ix)=(N0s(ix)*Bs-Cs)/(N0s(ix)*Bs+Cs);
        Ts(ix)=4*N0s(ix)*real(Nms(ix))/abs(N0s(ix)*Bs+Cs)^2;
        V_p = Mp(:,:,ix)*[1; Nmp(ix)]; Bp=V_p(1); Cp=V_p(2);
        rp(ix)=(N0p(ix)*Bp-Cp)/(N0p(ix)*Bp+Cp);
        Tp(ix)=4*N0p(ix)*real(Nmp(ix))/abs(N0p(ix)*Bp+Cp)^2;
    end
end
```

## 3. SCRIPT FOR EXAMPLE 1

```
% example1.m

% defining the wavelength range
nlamb=31; x=linspace(400,700,nlamb);

% defining the physical configuration
layers = {Inf 'Constant' {1 0};
           93 'Cauchy'   {1.36 4100. 0. 0. 0.};
          121 'Cauchy'   {2.0 17500. 98000. 0. 0.};
          185 'Cauchy'   {1.36 4100. 0. 0. 0.};
          Inf 'BK7'      {0} };
```

```
% defining the incidence angle
incang=0*pi/180 * linspace(1,1,length(x));

% calculate the overall coefficients
[rs,rp,Ts,Tp]=calc_rsrpTsTp(incang,layers,x);
R0=(abs(rs)).^2;        % Rp=(abs(rp)).^2; not needed since Rp=Rs at 0 deg

% calculate same for 40º incidence angle
incang=40*pi/180 * linspace(1,1,length(x));
[rs,rp,Ts,Tp]=calc_rsrpTsTp(incang,layers,x);
Rs40=(abs(rs)).^2;
Rp40=(abs(rp)).^2;

% calculate for configuration without layers at 0º incidence angle
NO_layers = {Inf 'Constant' {1. 0.};
             Inf 'BK7' {0} };
incang=0*pi/180 * linspace(1,1,length(x));
[rs,rp,Ts,Tp]=calc_rsrpTsTp(incang,NO_layers,x);
Rglass=(abs(rs)).^2;

% plot corresponding to Fig 3 in the text
figure, plot(x,R0,'k',x,Rs40,'k--',x,Rp40,'k-.',x,Rglass,'k+')
    legend('R0','Rs40','Rp40','Rglass')

% 30º linearly polarized light (for 40º incidence angle)
ang_fi=30*pi/180;       % vibration direction for linearly polarized light
R40P30=(Rs40*(sin(ang_fi))^2+Rp40*(cos(ang_fi))^2)/2; % reflected intensity

% non-polarized light (for 40º incidence angle)
R40Nat=(Rs40+Rp40)/2;   % reflected intensity for natural light

% plot corresponding to Fig 4 in the text
figure, plot(x,R40P30,'k-.',x,R40Nat,'k'), legend('R40P30','R40Nat')

% end of script Example1
```

## 4. SCRIPT FOR EXAMPLE 2

```
% example2.m

% defining the wavelength range
nlamb=31; x=linspace(400,700,nlamb);

% defining the physical configuration
layers_a = {Inf 'Constant'  {1. 0.};    % coating of the front (a) side
            200 'Sellmeier' {1.7 10000.};
             30 'Cauchy' {1.5 10000. 0. 0.1 150};
            Inf 'BK7'    {0}};

layers_b = {Inf 'BK7'    {0};            % coating of the back (b) side
            300 'Cauchy' {1.8 10000. 0. 0. 0.};
             30 'File'   {'EMA3_n_k.dat'};
            Inf 'Constant'  {1. 0.}};

% defining the 10º incidence angle
incang=10*pi/180*linspace(1,1,length(x));

% calculate the overall coeficients, 'a' side, Plus quantities
[rs,rp,Ts,Tp]=calc_rsrpTsTp(incang,layers_a,x);
Rs_aPlus=(abs(rs)).^2;
Rp_aPlus=(abs(rp)).^2;
```

```
    Ts_aPlus=Ts;
    Tp_aPlus=Tp;

    layers_ainv=layers_a(end:-1:1,:);        % inverts layers
    N0=calc_Nlayer(layers_a,x,1);        % ambient index, forward direction
    N0_inv=calc_Nlayer(layers_ainv,x,1);       % ambient index, backward direction
    incang_inv=asin(N0.*sin(incang)./N0_inv); % complex Snell law

    % calculate the overall coefficients, 'a' side, Minus quantities
    [rs,rp,Ts,Tp]=calc_rsrpTsTp(incang_inv,layers_ainv,x);
    Rs_aMinus=(abs(rs)).^2;
    Rp_aMinus=(abs(rp)).^2;
    Ts_aMinus=Ts;
    Tp_aMinus=Tp;

    % calculate the overall coefficients, 'b' side, Plus quantities
    [rs,rp,Ts,Tp]=calc_rsrpTsTp(incang_inv,layers_b,x);
    Rs_bPlus=(abs(rs)).^2;
    Rp_bPlus=(abs(rp)).^2;
    Ts_bPlus=Ts;
    Tp_bPlus=Tp;

    % calculate the overall R and T
    Rs=(Rs_aPlus+Rs_bPlus.*(Ts_aPlus.^2-Rs_aMinus.*Rs_bPlus)) ./ (1-Rs_aMinus.*Rs_bPlus);
    Rp=(Rp_aPlus+Rp_bPlus.*(Tp_aPlus.^2-Rp_aMinus.*Rp_bPlus)) ./ (1-Rp_aMinus.*Rp_bPlus);
    Ts=(Ts_aPlus.*Ts_bPlus) ./ (1-Rs_aMinus.*Rs_bPlus);
    Tp=(Tp_aPlus.*Tp_bPlus) ./ (1-Rp_aMinus.*Rp_bPlus);

    % plot corresponding to Fig 5 in the text
    figure, plot(x,Rs,'k.-',x,Rs_aPlus,'k:',x,Rp,'k--',x,Rp_aPlus,'k-.')
                    legend('Rs','Rsa','Rp','Rpa')

    % plot corresponding to Fig 6 in the text
    figure,
    plot(x,Ts_aPlus,'k.',x,Ts_aMinus,'k:',x,Tp_aPlus,'k--',x,Tp_aMinus,'k-.')
                    legend('TsaP','TsaM','TpaP','TpaM')

    % end of script example 2
```

## 5. SCRIPT FOR EXAMPLE 3

```
    % example3.m

    % defining the physical configuration
    layers =  {Inf 'Constant' {1.5 0.};
                30 'Constant' {0.056206 4.2776};
               Inf 'Constant' {1. 0.}};

    % defining the wavelength in nanometers
    x = 633.0;

    % defining 1000 incidence angles between 0º and 90º
    numang = 1000;
    Rs=zeros(1,numang); Rp=zeros(1,numang);
    angles = linspace(0,90,numang)*pi/180;

    % calculate Rs and Rp of Figure 7
    cont=0;
    for incang = angles
        cont=cont+1;
        [rs,rp,Ts,Tp]=calc_rsrpTsTp(incang,layers,x);
```

```
        Rs(cont)=abs(rs)^2; Rp(cont)=abs(rp)^2;
end
figure, plot(angles*180/pi,Rs,'k--',angles*180/pi,Rp,'k');
xlabel('incidence angle (deg)'), ylabel('Reflectance'), legend('Rs','Rp')

% defining 150 incidence angles and 100 thicknesses
numang = 150; angles = linspace(35,50,numang)*pi/180;
numthk = 100; thicks = linspace(0,100,numthk);
Rp=zeros(numthk,numang);

% calculate Rp of Figure 8
cont=0;
for incang = angles
    cont=cont+1;
    for ithk = 1:numthk
        layers{2,1}=thicks(ithk);
        [rs,rp,Ts,Tp]=calc_rsrpTsTp(incang,layers,x);
        Rp(ithk,cont)=abs(rp)^2;
    end
end
figure, pcolor(angles*180/pi,thicks,Rp), colormap gray
xlabel('incidence angle (deg)'),ylabel('thickness'), colorbar, shading interp

% definint configuration of Figure 9 (thickness 50 nm)
layers =  {Inf 'Constant' {1.5 0.};
            50 'Constant' {0.056206 4.2776};
           Inf 'Constant' {1 0};};

% calculate Rp of Figure 9
numang = 200; angles = linspace(40,90,numang)*pi/180;
numnout = 100; nout = linspace(1.0,1.5,numnout);
Rp = zeros(numnout,numang);
cont=0;
for incang = angles
    cont=cont+1;
    for inou=1:numnout
        layers{3,3} = {nout(inou),0};
        [rs,rp,Ts,Tp]=calc_rsrpTsTp(incang,layers,x);
        Rp(inou,cont) = abs(rp)^2;
    end
end
figure, pcolor(angles*180/pi,nout,Rp), colormap gray
xlabel('incidence angle (deg)'),ylabel('air side refractive index'), colorbar, shading interp;

% end of script example 3
```

## 6. USING THE EFFECTIVE MEDIUM APPROXIMATION

For an array of wavelengths 'x', the script creates a file 'EMA3_n_k.dat' containing the refractive index data computed by an EMA model from the porous mixture of two materials (Ag and $SiO_2$) whose refractive indices are stored in the two files 'Ag_n_k.dat' and 'SiO2_n.dat'. The porous mixture is modelled by the two materials plus 'Vacuum' (with refractive index equal to 1). The volume fractions of the three constitutive materials are named in the script, respectively, fr(1), fr(2) and fr(3).
For better illustration of how to read files and on how to manage minor reading variations in the script, the input data are effectively read differently for the two materials.
For 'Ag' the data are read from the 'nm-n-k' three-column file format 'Ag_n_k.dat' presented below. We always assume that the 'k' values we read and write in files are positive. To be consistent with our theoretical approach.
For 'SiO2' only the 'nm-n' data are read from the file 'SiO2_n.dat', because it is known in advance that all 'k' values are '0'. For the 'Vacuum', the index is in fact not read and simply is assumed to be a constant value '1'.
After running the script, it creates the file 'EMA3_n_k.dat', as shown below.

```
% fit_EMA.m
```

```
global N fr ct
nlamb=48; x=linspace(320,790,nlamb);   % check wavelength units!!
S1=load('Ag_n_k.dat');                 % read n k data
S2=load('SiO2_n.dat');                 % read only n data
S3=linspace(1.,1.,nlamb);              % impose index of vacuum=1.
fr(1)=.3;                              % volume fraction of material 1
N(1,:)=interp1(S1(:,1),S1(:,2)+1i*abs(S1(:,3)),x); % interpolates to 'x'
fr(2)=.6;                              % volume fraction of material 2
N(2,:)=interp1(S2(:,1),S2(:,2),x);     % interpolates to our values 'x'
fr(3)=.1;                              % volume fraction of material 3
N(3,:)=S3;                             % impose index of vacuum=1.
Vini=[real(sum(fr*N(:,1))),imag(sum(fr*N(:,1)))]; % starting point for search
for ct=1:nlamb                         % for each value in 'x'
   Vfin=fminsearch(@fmerit_EMA,Vini);  % fits according to EMA
   NN(1,ct)=Vfin(1); NN(2,ct)=Vfin(2);
   Vini=Vfin;                          % starting point for new search
   ct=ct+1;                            % counter
end
res=fopen('EMA3_n_k.dat','w');         % opens file for writing
fprintf(res,'%6.3f %6.3f %6.3f\n',[x;NN]);  % writes 3 columns
fclose(res);                           % closes file

function sumabs = fmerit_EMA(V)        % function for fitting
global N fr ct
sumabs=abs(sum(fr(:).*(N(:,ct).^2- ... (V(1)+1i*V(2))^2)./(N(:,ct).^2+2*(V(1)+1i*V(2))^2)));
end

% end of script fit_EMA
```

## A. FILE 'Ag_n_k.dat'

```
187.9   1.07   1.212
191.6   1.1    1.232
195.3   1.12   1.255
199.3   1.14   1.277
203.3   1.15   1.296
207.3   1.18   1.312
211.9   1.2    1.325
216.4   1.22   1.336
221.4   1.25   1.342
226.2   1.26   1.344
231.3   1.28   1.357
237.1   1.28   1.367
242.6   1.3    1.378
249.    1.31   1.389
255.1   1.33   1.393
261.6   1.35   1.387
268.9   1.38   1.372
276.1   1.41   1.331
284.4   1.41   1.264
292.4   1.39   1.161
300.9   1.34   0.964
310.7   1.13   0.616
320.4   0.81   0.392
331.5   0.17   0.829
342.5   0.14   1.142
354.2   0.1    1.419
367.9   0.07   1.657
381.5   0.05   1.864
```

| | | |
|---|---|---|
| 397.4 | 0.05 | 2.07 |
| 413.3 | 0.05 | 2.275 |
| 430.5 | 0.04 | 2.462 |
| 450.9 | 0.04 | 2.657 |
| 471.4 | 0.05 | 2.869 |
| 495.9 | 0.05 | 3.093 |
| 520.9 | 0.05 | 3.324 |
| 548.6 | 0.06 | 3.586 |
| 582.1 | 0.05 | 3.858 |
| 616.8 | 0.06 | 4.152 |
| 659.5 | 0.05 | 4.483 |
| 704.5 | 0.04 | 4.838 |
| 756. | 0.03 | 5.242 |
| 821.1 | 0.04 | 5.727 |
| 892. | 0.04 | 6.312 |
| 984. | 0.04 | 6.992 |
| 1088. | 0.04 | 7.795 |
| 1216. | 0.09 | 8.828 |
| 1393. | 0.13 | 10.1 |
| 1610. | 0.15 | 11.85 |
| 1937. | 0.24 | 14.08 |

**B.  FILE 'SiO2_n.dat'**

| | |
|---|---|
| 280 | 1.505961 |
| 285 | 1.504421 |
| 290 | 1.502966 |
| 295 | 1.501590 |
| 300 | 1.500288 |
| 305 | 1.499055 |
| 310 | 1.497885 |
| 315 | 1.496774 |
| 320 | 1.495719 |
| 325 | 1.494715 |
| 330 | 1.493759 |
| 335 | 1.492849 |
| 340 | 1.491981 |
| 345 | 1.491152 |
| 350 | 1.490361 |
| 355 | 1.489605 |
| 360 | 1.488882 |
| 365 | 1.488189 |
| 370 | 1.487527 |
| 375 | 1.486891 |
| 380 | 1.486282 |
| 385 | 1.485698 |
| 390 | 1.485137 |
| 395 | 1.484598 |
| 400 | 1.484080 |
| 405 | 1.483582 |
| 410 | 1.483102 |
| 415 | 1.482641 |
| 420 | 1.482197 |
| 425 | 1.481768 |
| 430 | 1.481356 |
| 435 | 1.480958 |
| 440 | 1.480573 |
| 445 | 1.480203 |
| 450 | 1.479844 |
| 455 | 1.479498 |
| 460 | 1.479164 |
| 465 | 1.478841 |
| 470 | 1.478528 |
| 475 | 1.478225 |
| 480 | 1.477932 |
| 485 | 1.477649 |
| 490 | 1.477374 |
| 495 | 1.477107 |
| 500 | 1.476849 |
| 505 | 1.476599 |
| 510 | 1.476356 |
| 515 | 1.476121 |
| 520 | 1.475892 |
| 525 | 1.475670 |
| 530 | 1.475455 |
| 535 | 1.475245 |
| 540 | 1.475042 |
| 545 | 1.474844 |
| 550 | 1.474652 |
| 555 | 1.474464 |
| 560 | 1.474283 |
| 565 | 1.474106 |
| 570 | 1.473933 |
| 575 | 1.473766 |
| 580 | 1.473602 |
| 585 | 1.473443 |
| 590 | 1.473288 |
| 595 | 1.473137 |
| 600 | 1.472990 |
| 605 | 1.472847 |
| 610 | 1.472707 |
| 615 | 1.472570 |
| 620 | 1.472437 |
| 625 | 1.472307 |
| 630 | 1.472181 |
| 635 | 1.472057 |
| 640 | 1.471936 |
| 645 | 1.471818 |
| 650 | 1.471703 |
| 655 | 1.471591 |
| 660 | 1.471481 |
| 665 | 1.471373 |
| 670 | 1.471268 |
| 675 | 1.471166 |
| 680 | 1.471065 |
| 685 | 1.470967 |
| 690 | 1.470871 |
| 695 | 1.470778 |
| 700 | 1.470686 |
| 705 | 1.470596 |
| 710 | 1.470508 |
| 715 | 1.470422 |
| 720 | 1.470338 |
| 725 | 1.470255 |
| 730 | 1.470174 |
| 735 | 1.470095 |
| 740 | 1.470018 |
| 745 | 1.469942 |
| 750 | 1.469867 |
| 755 | 1.469794 |
| 760 | 1.469723 |
| 765 | 1.469653 |
| 770 | 1.469584 |
| 775 | 1.469517 |
| 780 | 1.469451 |
| 785 | 1.469386 |
| 790 | 1.469323 |
| 795 | 1.469260 |
| 800 | 1.469199 |
| 805 | 1.469139 |
| 810 | 1.469080 |
| 815 | 1.469023 |
| 820 | 1.468966 |
| 825 | 1.468910 |
| 830 | 1.468855 |

## C. FILE 'EMA3_n_k.dat'

| | | |
|---|---|---|
| 320.000 | 1.234 | 0.131 |
| 330.000 | 0.991 | 0.287 |
| 340.000 | 0.979 | 0.464 |
| 350.000 | 1.023 | 0.579 |
| 360.000 | 1.062 | 0.659 |
| 370.000 | 1.096 | 0.720 |
| 380.000 | 1.128 | 0.770 |
| 390.000 | 1.159 | 0.809 |
| 400.000 | 1.188 | 0.844 |
| 410.000 | 1.218 | 0.877 |
| 420.000 | 1.242 | 0.907 |
| 430.000 | 1.265 | 0.934 |
| 440.000 | 1.286 | 0.956 |
| 450.000 | 1.306 | 0.977 |
| 460.000 | 1.329 | 0.999 |
| 470.000 | 1.352 | 1.020 |
| 480.000 | 1.371 | 1.039 |
| 490.000 | 1.390 | 1.057 |
| 500.000 | 1.408 | 1.075 |
| 510.000 | 1.427 | 1.093 |
| 520.000 | 1.446 | 1.110 |
| 530.000 | 1.465 | 1.127 |
| 540.000 | 1.484 | 1.143 |
| 550.000 | 1.503 | 1.159 |
| 560.000 | 1.518 | 1.174 |
| 570.000 | 1.533 | 1.188 |
| 580.000 | 1.548 | 1.202 |
| 590.000 | 1.564 | 1.216 |
| 600.000 | 1.580 | 1.229 |
| 610.000 | 1.597 | 1.242 |
| 620.000 | 1.612 | 1.255 |
| 630.000 | 1.626 | 1.267 |
| 640.000 | 1.640 | 1.279 |
| 650.000 | 1.654 | 1.292 |
| 660.000 | 1.668 | 1.303 |
| 670.000 | 1.681 | 1.315 |
| 680.000 | 1.695 | 1.327 |
| 690.000 | 1.709 | 1.339 |
| 700.000 | 1.723 | 1.350 |
| 710.000 | 1.737 | 1.361 |
| 720.000 | 1.750 | 1.372 |
| 730.000 | 1.764 | 1.383 |
| 740.000 | 1.777 | 1.394 |
| 750.000 | 1.791 | 1.405 |
| 760.000 | 1.804 | 1.415 |
| 770.000 | 1.817 | 1.424 |
| 780.000 | 1.830 | 1.433 |
| 790.000 | 1.843 | 1.442 |

## 7. SCRIPTS IN PYTHON

Please note that the codes below mimic the Matlab codes explained in the text. They lack Python programming style but they have been tested to work.

```
# Example1.py

# -*- coding: utf-8 -*-
from numpy import *
from matplotlib.pyplot import *
from Funcs import *

nlamb=31; x=linspace(400,700,nlamb);
layers = [ [nan, "Constant", [1.,0.]],
           [ 93, "Cauchy", [1.36,4100.,0.,0.,0.]],
           [121, "Cauchy", [2.0,17500.,98000.,0.,0.]],
           [185, "Cauchy", [1.36,4100.,0.,0.,0.]],
           [nan, "BK7", [0]] ];
incang=0*pi/180*ones(x.size)
[rs,rp,Ts,Tp]=calc_rsrpTsTp(incang,layers,x);
R0=(abs(rs))**2;

incang=40*pi/180*ones(x.size)
[rs,rp,Ts,Tp]=calc_rsrpTsTp(incang,layers,x);
Rs40=(abs(rs))**2; Rp40=(abs(rp))**2;

NO_layers = [ [nan, "Constant", [1.,0.]],
              [nan, "BK7", [0]] ]
incang=0*pi/180*ones(x.size)
[rs,rp,Ts,Tp]=calc_rsrpTsTp(incang,NO_layers,x);
Rglass=(abs(rs))**2;
figure(), plot(x,R0,label='R0'), plot(x,Rs40,label='Rs40')
plot(x,Rp40,label='Rp40'), plot(x,Rglass,label='Rglass'), legend()

ang_fi=30.0*pi/180
R40P30=(Rs40*(sin(ang_fi))**2+Rp40*(cos(ang_fi))**2)/2;
R40Nat=(Rs40+Rp40)/2;

figure(), plot(x,R40P30,label='R40P30'), plot(x,R40Nat,label='R40Nat'), legend()

# Example2.py

# -*- coding: utf-8 -*-
from numpy import *
from matplotlib.pyplot import *
from Funcs import *

nlamb=31; x=linspace(400,700,nlamb)
layers_a = [ [nan, 'Constant', [1.,0.]],
             [200, 'Sellmeier', [1.7,10000]],
             [ 30, 'Cauchy', [1.5,10000.,0.,0.1,150.]],
             [nan, 'BK7', [0]] ]

layers_b = [ [nan, 'BK7', [0]],
             [300, 'Cauchy', [1.8,10000.,0.,0.,0.]],
             [ 30, 'File', ['EMA3_n_k.dat']],
             [nan, 'Constant', [1.,0.]] ]

incang=10*pi/180*ones(x.size)
[rs,rp,Ts,Tp]=calc_rsrpTsTp(incang,layers_a,x)
Rs_aPlus=(abs(rs))**2; Rp_aPlus=(abs(rp))**2; Ts_aPlus=Ts; Tp_aPlus=Tp
```

```python
layers_ainv=layers_a[::-1]
N0=calc_Nlayer(layers_a,x,0)
N0_inv=calc_Nlayer(layers_ainv,x,0)
incang_inv=arcsin(N0*sin(incang)/N0_inv)

[rs,rp,Ts,Tp]=calc_rsrpTsTp(incang_inv,layers_ainv,x)
Rs_aMinus=(abs(rs))**2; Rp_aMinus=(abs(rp))**2; Ts_aMinus=Ts; Tp_aMinus=Tp

[rs,rp,Ts,Tp]=calc_rsrpTsTp(incang_inv,layers_b,x)
Rs_bPlus=(abs(rs))**2; Rp_bPlus=(abs(rp))**2; Ts_bPlus=Ts; Tp_bPlus=Tp

Rs=(Rs_aPlus+Rs_bPlus*(Ts_aPlus**2-Rs_aMinus*Rs_bPlus)) / (1-Rs_aMinus*Rs_bPlus)
Rp=(Rp_aPlus+Rp_bPlus*(Tp_aPlus**2-Rp_aMinus*Rp_bPlus)) / (1-Rp_aMinus*Rp_bPlus)
Ts=(Ts_aPlus*Ts_bPlus) / (1-Rs_aMinus*Rs_bPlus)
Tp=(Tp_aPlus*Tp_bPlus) / (1-Rp_aMinus*Rp_bPlus)

figure(), plot(x,Rs,'k',label='Rs'), plot(x,Rs_aPlus,'k:',label='Rsa')
plot(x,Rp,'r',label='Rp'), plot(x,Rp_aPlus,'r:',label='Rpa'), legend()
figure(), plot(x,Ts_aPlus,label='TsaP'), plot(x,Ts_aMinus,label='TsaM')
plot(x,Tp_aPlus,label='TpaP'), plot(x,Tp_aMinus,label='TpaM'), legend()

# Example3.py

# -*- coding: utf-8 -*-
from numpy import *
from matplotlib.pyplot import *
from Funcs import *

layers = [ [nan, "Constant", [1.5,0.]],
           [ 30, "Constant", [0.056206,4.2776]],
           [nan, "Constant", [1.,0.]] ]
x = linspace(633.,633,1)
numang = 1000
Rs=zeros(numang,dtype=cfloat); Rp=zeros(numang,dtype=cfloat)
angles = linspace(0,90,numang)*pi/180;
cont=0
for incang in angles:
    [rs,rp,Ts,Tp]=calc_rsrpTsTp(incang,layers,x)
    # Rs(cont)=abs(rs)**2; Rp(cont)=abs(rp)**2
    Rs[cont]=abs(rs)**2; Rp[cont]=abs(rp)**2
    cont=cont+1
figure(); plot(angles*180/pi,Rs,label='Rs'); plot(angles*180/pi,Rp,label='Rp')
xlabel('incidence angle (deg)'); ylabel('Reflectance'); legend()

numang = 150; angles = linspace(35,50,numang)*pi/180
numthk = 100; thicks = linspace(0,100,numthk)
Rp=zeros([numthk,numang],dtype=float)
cont=0
for incang in angles:
    for ithk in range(0,numthk):
        layers[1][0]=thicks[ithk]
        [rs,rp,Ts,Tp]=calc_rsrpTsTp(incang,layers,x)
        Rp[ithk,cont]=abs(rp)**2
    cont=cont+1
figure(); c=pcolor(angles*180/pi,thicks,Rp,cmap='jet')
colorbar(c,orientation='vertical')
xlabel('incidence angle (deg)'), ylabel('thickness')

layers = [ [nan, "Constant", [1.5,0.]],
           [ 50, "Constant", [0.056206,4.2776]],
           [nan, "Constant", [1.,0.]] ]
numang = 200; angles = linspace(40,90,numang)*pi/180;
```

```python
numnout = 100; nout = linspace(1.0,1.5,numnout);
Rp = zeros([numnout,numang],dtype=float)
cont=0;
for incang in angles:
    for inou in range(0,numnout):
        layers[2][2][0] = nout[inou]
        [rs,rp,Ts,Tp]=calc_rsrpTsTp(incang,layers,x);
        Rp[inou,cont] = abs(rp)**2
    cont=cont+1
figure(); c=pcolor(angles*180/pi,nout,Rp,cmap='jet')
colorbar(c,orientation='vertical')
xlabel('incidence angle (deg)'),ylabel('air side refractive index')

# Funcs.py

# -*- coding: utf-8 -*-
from numpy import *

def calc_Nlayer(layers,x,num_lay):
    case = layers[num_lay][1]
    if case == 'Constant':
        v2p=[layers[num_lay][2][0],layers[num_lay][2][1]]
        nnn=v2p[0]; kap=abs(v2p[1])
        Nlay=(nnn-1j*kap)*ones(x.size)
    elif case == 'Cauchy':
        v5p=[layers[num_lay][2][0],layers[num_lay][2][1],layers[num_lay][2][2],
            layers[num_lay][2][3],layers[num_lay][2][4]]
        nnn=v5p[0]+v5p[1]/x**2+v5p[2]/x**4;  kap=abs(v5p[3])*exp(v5p[4]/x)
        Nlay=nnn-1j*kap
    elif case == 'Sellmeier':
        v2p=[layers[num_lay][2][0],layers[num_lay][2][1]]
        nnn=sqrt(1+v2p[0]*x**2/(x**2-v2p[1]));   kap=0.
        Nlay=nnn-1j*kap
    elif case == 'File':
        aux=loadtxt(layers[num_lay][2][0])  # N,k data
        nnn=interp(x,aux[:,0],aux[:,1])
        kap=interp(x,aux[:,0],aux[:,2])
        Nlay=nnn-1j*abs(kap);
    elif case == 'BK7':
        n2=1+(1.03961*x**2)/(x**2-6.0e3)+(0.23179*x**2)/ \
          (x**2-2.0e4)+(1.0146*x**2)/(x**2-1.0e8)
        nnn=sqrt(n2)
        Nlay=nnn-1j*0.0
    return Nlay

def calc_rsrpTsTp(incang,layers,x):
    Ms=zeros([x.size,2,2],dtype=cfloat); Mp=zeros([x.size,2,2],dtype=cfloat)
    S=zeros([x.size,2,2],dtype=cfloat); P=zeros([x.size,2,2],dtype=cfloat)
    Ms[:,0,0]=1; Ms[:,1,1]=1; Mp[:,0,0]=1; Mp[:,1,1]=1
    rs=zeros((x.size),dtype=cfloat); rp=zeros((x.size),dtype=cfloat)
    Ts=zeros((x.size),dtype=cfloat); Tp=zeros((x.size),dtype=cfloat);
    im=0
    N0=calc_Nlayer(layers,x,im)
    N0s=N0*cos(incang); N0p=N0/cos(incang)
    for im in range(1,len(layers)-1):
        Nlay=calc_Nlayer(layers,x,im)
        ARR=sqrt(Nlay**2-N0**2*(sin(incang)**2))
        Ns=abs(real(ARR)) - 1j*abs(imag(ARR))
        d=layers[im][0]
        Dr=2*pi*d/x*Ns
        Np=Nlay**2/Ns
        for ix in range(x.size):
```

```
            S[ix,:,:]=[[cos(Dr[ix]), 1j/Ns[ix]*sin(Dr[ix])],
             [1j*Ns[ix]*sin(Dr[ix]), cos(Dr[ix])]];
            P[ix,:,:]=[[cos(Dr[ix]), 1j/Np[ix]*sin(Dr[ix])],
             [1j*Np[ix]*sin(Dr[ix]), cos(Dr[ix])]];
            Ms[ix,:,:]=Ms[ix,:,:]@S[ix,:,:]; Mp[ix,:,:]=Mp[ix,:,:]@P[ix,:,:]
    im=len(layers)-1
    Nm=calc_Nlayer(layers,x,im)
    ARR=sqrt(Nm**2-N0**2*(sin(incang)**2))
    Nms=abs(real(ARR)) - 1j*abs(imag(ARR))
    Nmp=Nm**2/Nms
    for ix in range(x.size):
        V_s = Ms[ix,:,:]@[[1.],[Nms[ix]]]; Bs=V_s[0]; Cs=V_s[1]
        rs[ix]=(N0s[ix]*Bs-Cs)/(N0s[ix]*Bs+Cs)
        Ts[ix]=4*N0s[ix]*real(Nms[ix])/(abs(N0s[ix]*Bs+Cs))**2
        V_p = Mp[ix,:,:]@[[1.],[Nmp[ix]]]; Bp=V_p[0]; Cp=V_p[1]
        rp[ix]=(N0p[ix]*Bp-Cp)/(N0p[ix]*Bp+Cp)
        Tp[ix]=4*N0p[ix]*real(Nmp[ix])/(abs(N0p[ix]*Bp+Cp))**2
    return [rs,rp,Ts,Tp]
```